\documentclass[journal=jctcce,manuscript=article]{achemso}
\setkeys{acs}{maxauthors=6,etalmode=truncate,chaptertitle =true,articletitle=true}
\newcommand{\onlinecite}[1]{\hspace{-1 ex} \nocite{#1}\citenum{#1}} 
\usepackage[version=3]{mhchem} 
\usepackage[T1]{fontenc}       
\usepackage{epstopdf}
\usepackage{graphicx}
\usepackage{longtable}
\usepackage{float}
\usepackage{color}
\usepackage[none]{hyphenat}
\usepackage[dvipsnames]{xcolor}
\title
{How accurate is density functional theory at predicting dipole moments? An assessment using a new database of 200 benchmark values.}
\author{Diptarka Hait}
\affiliation
{{Kenneth S. Pitzer Center for Theoretical Chemistry, Department of Chemistry, University of California, Berkeley, California 94720, USA}}

\author{Martin Head-Gordon}
\email{mhg@cchem.berkeley.edu}
\affiliation
{{Kenneth S. Pitzer Center for Theoretical Chemistry, Department of Chemistry, University of California, Berkeley, California 94720, USA}}
\affiliation{Chemical Sciences Division, Lawrence Berkeley National Laboratory, Berkeley, California 94720, USA}

\begin{document}
\begin{abstract}
Dipole moments are a simple, global measure of the accuracy of the electron density of a polar molecule. Dipole moments also affect the interactions of a molecule with other molecules as well as electric fields. To directly assess the accuracy of modern density functionals for calculating dipole moments, we have developed a database of 200 benchmark dipole moments, using coupled cluster theory through triple excitations, extrapolated to the complete basis set limit. This new database is used to assess the performance of {88 popular or recently developed} density functionals.  {The results suggest that double hybrid functionals perform the best, yielding dipole moments within about 3.6-4.5\% regularized RMS error versus the reference values---which is not very different from the 4\% regularized RMS error produced by coupled cluster singles and doubles. Many hybrid functionals also perform quite well, generating regularized RMS errors in the 5-6\% range. Some functionals however exhibit large outliers and local functionals in general perform less well than hybrids or double hybrids.}
 




\end{abstract}

\begin{tocentry}
%
%
%
%
%
\centering	\includegraphics[width=8.9cm, height=3.6cm]{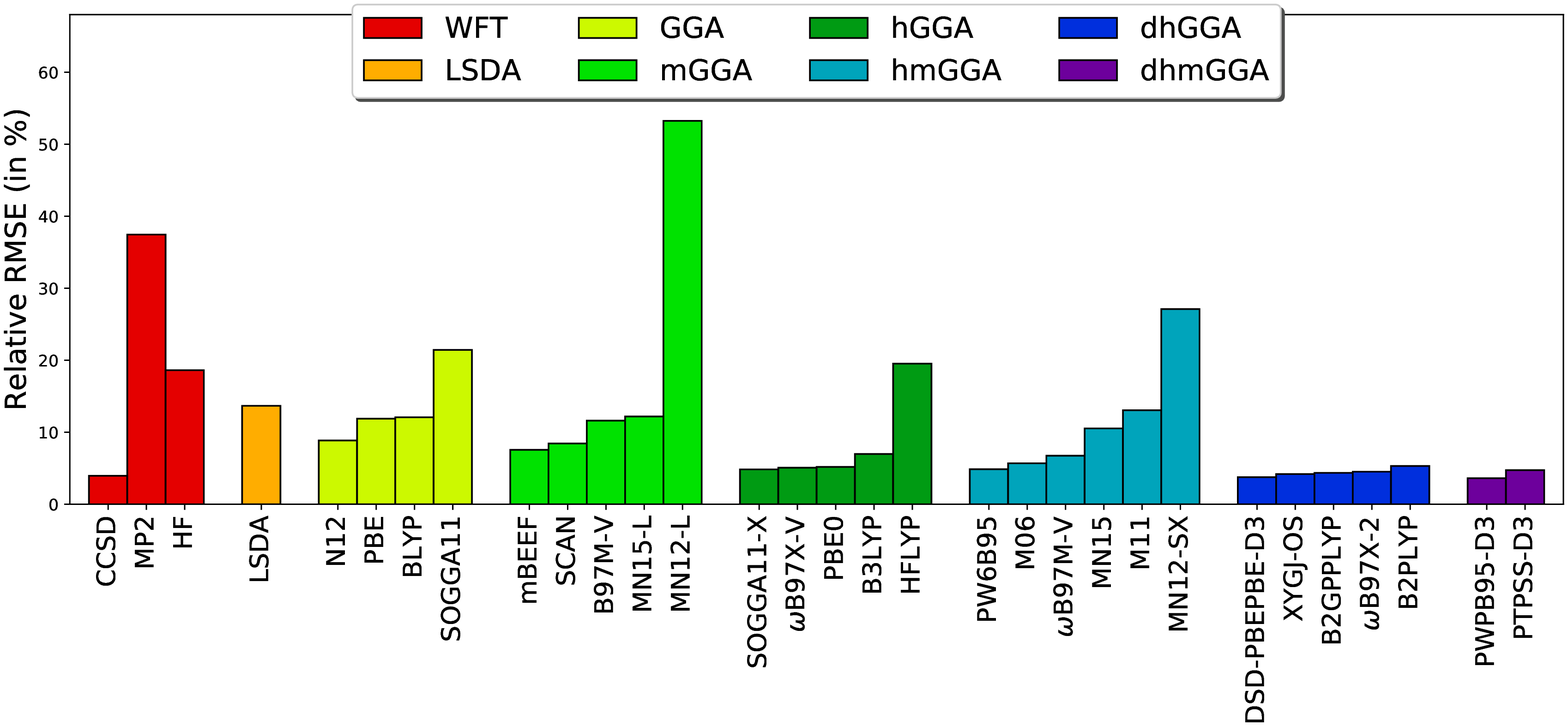}
\end{tocentry}





\maketitle
\section{Introduction}
Density functional theory (DFT) is currently the most popular approach for calculating the electronic structure of molecules and extended materials\cite{becke2014perspective,jones2015density,mardirossian2017thirty}. Although DFT is formally exact\cite{hohenberg1964inhomogeneous}, the true functional that maps electron density to electronic energy remains unknown. Most practical calculations thus employ various density functional approximations (DFAs) within the Kohn--Sham\cite{kohn1965self} DFT framework. Hundreds of DFAs have consequently been proposed over the last few decades, with development strategies focusing on empirical fitting to large benchmark datasets\cite{m06,m06l,MN15}, satisfaction of various exact physical constraints\cite{Slater,VWN,PBE} or intermediate routes combining aspects of both approaches\cite{wB97MV,SCAN}. With rare exceptions\cite{b97-2,HCTH,hcth120,HCTH3}, fitting based approaches have historically favored recovering accurate energies over {accurate} densities, in the hope that DFAs predicting improved energies were better approximations to the exact functional, and would thus automatically yield more accurate electron densities.  

This approach has recently been questioned by studies that compared coupled cluster singles and doubles (CCSD) electron densities of a few atomic\cite{medvedev2017density} and diatomic\cite{brorsen2017accuracy} species to those predicted by various DFAs. These studies appear to suggest that many modern functionals (that are excellent for energetics) are unable to predict very good densities, deviating from an earlier trend of DFAs simultaneously improving energies and densities. This has led to suggestions that DFT has strayed from the path to the exact functional\cite{medvedev2017density}, causing lively discussions in the scientific community \cite{hammes2017conundrum,kepp2017density,medvedev2017response,gould2017makes,graziano2017quantum,korth2017density,verma2017can,wang2017well,kepp2017energy,ranasinghe2017note,mayer2017conceptual}.
Discussions about ``true paths'' aside, accurate electronic densities are necessary not only for intellectual completeness, but also to quantitatively predict the response of the electronic energy to external potentials (such as ones in spectroscopic experiments, or intermolecular interactions or at an electrode). 
It is therefore desirable to develop functionals that would simultaneously predict the energy and density to high levels of accuracy, perhaps by incorporating density information as part of functional development.

Molecular multipole moments describe spatial moments of the electron density, and the accuracy with which they can be calculated is an integrated measure of the quality of the corresponding density. The simplest possible multipole is the dipole $\vec{\mu}=\displaystyle\int \vec{r}\rho(\vec{r})d\vec{r}$ (where $\rho(\vec{r})$ is \textit{charge} density), which is the sum of the first spatial moment of the electron density and a nuclear charge contribution. $\vec{\mu}$ is not a perfect stand-in for the density (not least because it can vanish due to symmetry) and it is certainly possible for an incorrect density to accidentally yield a correct value of $\vec{\mu}$.  However, $\vec{\mu}$ is perhaps the simplest observable that captures errors in the underlying density. $\vec{\mu}$ also determines the response of the energy to a constant electric field and is relevant to long-range electrostatics in force field parametrization\cite{vanommeslaeghe2010charmm,ponder2003force,ponder2010current}, making it a relevant density derived quantity to examine for DFA testing and development. While we do not consider them here, higher moments like the quadrupole and higher response properties like polarizability are also relevant to the performance of DFAs. 

Although there have been previous studies examining the accuracy of DFT in estimating vibrationally averaged experimental dipole moments\cite{hickey2014benchmarking,verma2017can}, there does not appear to be any large dataset of benchmark equilibrium $\vec{\mu}$. The distinction between experimental and equilibrium dipoles is important as they are known to differ by 3\% or more in many cases like NH$_3$\cite{puzzarini2008ab}. The accuracy of electronic structure approximations (in the absence of vibrational corrections) in directly reproducing experimental dipole moments thus may not be indicative of the quality of the approximation. This led us to develop a dataset of 200 coupled cluster dipole moments for the purpose of assessing and developing density functionals. The main constituent of our dataset is an estimate of the unrestricted CCSD(T)\cite{raghavachari1989fifth} dipole moment at the complete basis set (CBS) limit for 95 {normal valent} systems (88 molecules and 7 non-covalent complexes, containing up to 3 non-hydrogen atoms {and with all atoms at normal valence}) and 57 incomplete valence and open-shell  systems (50 molecules and 7 non-covalent complexes {where not all atoms were at normal valence}) at their equilibrium geometries (given in the Supporting Information). For simplicity, we only report magnitudes $\mu=|\vec{\mu}|$ for most species, but we do supply the direction for certain cases like CO, where different DFAs predict different polarities. The remaining data-points are $\vec{\mu}$ along the potential energy surfaces (PES) of FH and FCl, enabling exploration of some non-equilibrium configurations. We also assessed the performance of {88 popular or recent} density functionals\cite{Slater,VWN,PW92,pz81,b97d,PBE,b88,lyp,vv10,revPBE,tpss,revTPSS,tpssh,revtpssh,pbe0,p86,pw91,gam,hcth120,n12,sogga,sogga11,sogga11x,n12sx,b3lyp,b97mv,b97mrv,wb97xv,wB97MV,wB97XD,wB97,wB97XD3,wM05D,m06,m05,m052x,m06hf,m06l,m08so,m11,m11l,sherrill1999performance,mn12l,mn15l,ms2,mvs,scan0,b97,b97-1,b97-2,bmk,camb3lyp,hsehjs,lrcwpbe,lrcwpbeh,pw6b95,thctch,MN15,SCAN,wellendorff2014mbeef,revm06l,verma2014increasing,jin2016qtp,rcamb3lyp,chai2009long,zhang2009doubly,zhang2011fast,grimme2006semiempirical,karton2008highly} in predicting $\mu$ of the 152 dataset species at equilibrium geometry, in order to gauge the relative accuracy of these functionals in predicting densities over a substantially larger dataset than those employed by previous studies\cite{medvedev2017density,brorsen2017accuracy}. 


\section{Computational Methods}
All the calculations were done using a development version of Q-Chem 4\cite{QCHEM4}. The equilibrium geometries employed to estimate $\vec{\mu}$ for each of the 152 species considered were obtained from a variety of sources and no geometry re-optimization was done for individual methods. Experimental equilibrium geometries (obtained from the NIST Computational Chemistry Database\cite{johnson2015nist} with two exceptions: BH$_2$F\cite{takeo1993microwave} and BH$_2$Cl\cite{kawashima1993microwave}) were employed whenever possible for the molecular species. Molecular geometries were optimized at the MP2/cc-pVTZ level of theory when experimental geometries were unavailable, { as MP2 equilibrium geometries in general are known to be sufficiently accurate\cite{curtiss1998gaussian}}.  The geometries of the non-covalently bounded complexes were taken from the NCCE31\cite{zhao2005benchmark} (for closed-shell species) and TA13\cite{tentscher2013binding} (for open-shell species) databases.  All the molecular geometries employed in the present study are provided in the Supporting Information .

{Hartree-Fock (HF)} $\vec{\mu}$ were obtained from spin unrestricted calculations with the aug-cc-pCV5Z\cite{dunning1989gaussian,woon1995gaussian,peterson2002accurate,prascher2011gaussian} basis, which appeared to  have essentially converged to the CBS limit as they only showed a 0.04\% RMS deviation versus the equivalent aug-cc-pCVQZ numbers. Stability analysis was performed at the aug-cc-pCVQZ level to ensure all solutions were at a minima.  Spin unrestricted DFT calculations were done with the aug-pc-4\cite{jensen2001polarization,jensen2002polarization,jensen2002polarizationiii,jensen2004polarization,jensen2007polarization} basis {for functionals spanning Rungs 1-4 in Jacob's Ladder} and it was assumed that such 5$\zeta$ basis results ought to be essentially at CBS as well (though this may not strictly be true for a few ill-behaved functionals, as discussed in the results section). Local xc integrals were calculated over a radial grid with 99 points and an angular Lebedev grid with 590 points for all atoms, while non-local VV10 correlation was calculated over an SG-1\cite{gill1993standard} grid {(which consists of a subset of points employed in a grid with 50 radial and 194 angular points)}. Stability analysis was done at the aug-pc-2 level to determine which SCF solutions were potentially unstable, and the problematic aug-pc-4 cases were reoptimized to ensure that they were at a minima.  

Dipole moments for correlated wave function methods (MP2/CCSD/CCSD(T)) were obtained from two point central finite differences (using an electric field strength of $10^{-4}$ a.u.). The CBS limit was obtained by extrapolating the correlation component of $\mu$ from aug-cc-pCV5Z and aug-cc-pCVQZ results for the smaller species; and from the  aug-cc-pCVQZ and aug-cc-pCVTZ results for the remainder of the dataset. The extrapolation was done via the two point formula $\mu^{\mbox{corr}}_n=\mu^{\mbox{corr}}_{\infty}+A/n^3$ used in Ref [\onlinecite{halkier1999basis}], which was found to be accurate to $\approx0.2\%$ in predicting aug-cc-pCV5Z dipoles of some of the smaller species from aug-cc-pCVQZ and aug-cc-pCVTZ results, indicating that it was reasonably accurate. All the calculations were {done using unrestricted orbitals} (unless specified otherwise) and no orbitals were held frozen. 

{All calculations with double hybrid functionals for species at equilibrium geometry were done with restricted orbitals on account of N-representability violations in unrestricted MP2\cite{Kurlancheek2009}. Dipole moments were calculated using the same finite difference approach as correlated wave function methods. The non-MP2 component of $\vec{\mu}$ was assumed to have essentially converged to CBS at the aug-cc-pCV5Z level, while the MP2 correlation component was extrapolated to CBS from aug-cc-pCVQZ and aug-cc-pCVTZ results, using the same formula as above. The xc integrals for double hybrid functionals were calculated using the same grids as all other functionals. Calculations employing the XYGJ-OS\cite{zhang2011fast} functional were accelerated using the RI approximation\cite{feyereisen1993use} with the riMP2-cc-pVTZ auxiliary basis\cite{weigend1998ri,weigend2002efficient} for aug-cc-pCVTZ calculations and riMP2-cc-pVQZ auxiliary basis\cite{weigend2002efficient} for aug-cc-pCVQZ calculations.}

Wave function theory (WFT) calculations for the FH and FCl PES's were performed in the same manner as {equilibrium calculations}, with the CBS limit for correlated methods found from two point extrapolation from aug-cc-pCVQZ and aug-cc-pCVTZ results. {Double hybrid calculations were performed similarly to equilibrium calculations as well, with the major differences being use of unrestricted orbitals (since restricted references are grossly inadequate in describing bond dissociations \cite{szabo2012modern}) and treatment of aug-cc-pCVQZ numbers as the CBS limit for the non-MP2 component instead of aug-cc-pCV5Z results.}
Other DFT calculations were done with the aug-pc-3 basis, with the same integration grids as the equilibrium cases. Stability analysis was performed on all SCF solutions to ensure the determinant was a minimum. No expensive aug-pc-4{/aug-cc-pCV5Z} calculations were performed for any of the functionals, as the basis set incompleteness errors at the {quadruple $\zeta$ level} are small compared to the qualitative questions that the exploration along those potential energy curves are addressing.

The error in $\mu$ against the reference value $\mu_{\mathrm{ref}}$ is defined to be $\dfrac{\mu-\mu_{\mathrm{ref}}}{\mathrm{max}\left(\mu_{\mathrm{ref}},1 \mbox{ D}\right)}\times 100\%$. This regularized form was chosen to ensure that the errors were relative for large $\mu_{\mathrm{ref}}$ (to prevent relatively small perturbations of density in ionic species from dominating) but were absolute for small $\mu_{\mathrm{ref}}$ (to prevent a too small denominator from skewing the analyis). This effectively results in a regularized relative error as a majority of the database has $\mu_{\mathrm{ref}}\ge 1$ D. The choice of $1$ D as the cutoff is somewhat arbitrary, but it allows a seamless transition from absolute error (for small $\mu_{\mathrm{ref}}$) to relative error (for large $\mu_{\mathrm{ref}}$). Purely absolute and purely relative errors are also supplied in the Supporting Information, with trends in the former being consistent with the trends in our regularized relative error. Purely relative errors are however considerably less useful, as species with $<0.1$ D (such as the OF radical) wind up heavily dominating total error and the rest barely make a significant contribution.
 
\section{Results for equilibrium geometries}
We divided the 152 species in our dataset into two groups: 81 where the stable HF solution did not have spin-polarization, and 71 where the stable HF solution was spin-polarized. The former subset is henceforth referred to as NSP (not spin-polarized, or in other words, spin-restricted) and the latter as SP (spin-polarized, or unrestricted). Conventional electronic structure methods like CCSD, MP2 or the various DFAs are unlikely to experience catastrophic failure over the NSP subset, but the SP species have the potential to be more challenging.  

\begin{table*}[htb!]
	{	\centering
		\setlength\tabcolsep{3pt}
		\scriptsize{	\begin{tabular}{llrrrrr|cllrrrrr}
				Method   & Class & \multicolumn{3}{c}{RMSE} & MAX    & ME    && Method    & Class & \multicolumn{3}{c}{RMSE} & MAX    & ME    \\
				&       & All    & NSP    & SP     &        &       &  &         &       & All    & NSP    & SP     &        &       \\
				CCSD      & WFT   & 3.95   & 3.01   & 4.8    & 12.87  & 1.86  &  & $\omega$B97X-D       & hGGA   & 5.36   & 4.43   & 6.25   & 20.8   & 1.35  \\
				MP2       & WFT   & 37.45  & 3.96   & 54.63  & 367.72 & 7.97  &  & camB3LYP      & hGGA   & 5.83   & 3.68   & 7.56   & 40.23  & 1.7   \\
				RMP2      & WFT   & 7.32   & 3.96   & 9.84   & 32.49  & 0.55  &  & B97           & hGGA   & 6.06   & 4.99   & 7.09   & 29.03  & -1.23 \\
				HF        & WFT   & 18.61  & 12.81  & 23.53  & 57.35  & 7.18  &  & CAM-QTP01     & hGGA   & 6.88   & 4.94   & 8.57   & 36.16  & 3.62  \\
				RHF       & WFT   & 20.1   & 12.81  & 26.03  & 76.64  & 9.67  &  & B3LYP         & hGGA   & 6.98   & 4.43   & 9.05   & 46     & -0.82 \\
				&       &        &        &        &        &       &  & rCAM-B3LYP    & hGGA   & 7.48   & 5.43   & 9.27   & 35.23  & 4.18  \\
				SVWN1RPA  & LSDA  & 13.67  & 8.93   & 17.59  & 66.22  & -2.13 &  & HFLYP         & hGGA   & 19.53  & 14.87  & 23.76  & 66.79  & 9.11  \\
				Slater    & LSDA  & 14.82  & 12.11  & 17.4   & 49.31  & -3.95 &  &               &        &        &        &        &        &       \\
				&       &        &        &        &        &       &  & PW6B95        & hmGGA  & 4.87   & 3.76   & 5.88   & 24.18  & -0.44 \\
				N12       & GGA   & 8.85   & 5.13   & 11.74  & 45.29  & -1.72 &  & M05           & hmGGA  & 4.95   & 4.51   & 5.4    & 18.47  & -0.14 \\
				B97-D     & GGA   & 9.87   & 7.95   & 11.68  & 43.3   & -3.48 &  & $\omega$M05-D        & hmGGA  & 4.96   & 4.46   & 5.49   & 17.96  & 0.87  \\
				PBE       & GGA   & 11.89  & 9.55   & 14.09  & 45.02  & -4.15 &  & M06           & hmGGA  & 5.69   & 5.69   & 5.68   & 19.27  & -0.63 \\
				BLYP      & GGA   & 12.08  & 8.55   & 15.13  & 55.82  & -4.16 &  & $\omega$B97M-V       & hmGGA  & 5.74   & 4.1    & 7.16   & 31.94  & 0.18  \\
				SOGGA11   & GGA   & 21.44  & 11.31  & 28.95  & 197.74 & -2.35 &  & SCAN0         & hmGGA  & 6.28   & 4.46   & 7.86   & 22.56  & 1.17  \\
				&       &        &        &        &        &       &  & MS2h          & hmGGA  & 7.08   & 6.13   & 8.02   & 24.82  & -1.58 \\
				mBEEF     & mGGA  & 7.56   & 6.55   & 8.56   & 29.97  & -2.45 &  & TPSSh         & hmGGA  & 7.42   & 5.96   & 8.79   & 31.94  & -1.63 \\
				SCAN      & mGGA  & 8.44   & 6.8    & 9.98   & 31.81  & -1.31 &  & M06-2X        & hmGGA  & 7.94   & 3.76   & 10.9   & 67.07  & 1     \\
				MS2       & mGGA  & 8.94   & 7.76   & 10.12  & 28.98  & -2.83 &  & MN15          & hmGGA  & 10.53  & 4.05   & 14.78  & 79.28  & 0.45  \\
				TPSS      & mGGA  & 9.93   & 7.92   & 11.82  & 40.51  & -3.02 &  & BMK           & hmGGA  & 11.1   & 3.59   & 15.77  & 125.58 & 1.51  \\
				revM06-L  & mGGA  & 10.68  & 10.86  & 10.46  & 34.1   & -1.81 &  & M11           & hmGGA  & 13.06  & 5.51   & 18.18  & 135.49 & 4.67  \\
				M11-L     & mGGA  & 11.05  & 11.01  & 11.09  & 34.73  & -5.65 &  & M06-HF        & hmGGA  & 18.13  & 13.15  & 22.5   & 110.82 & 6.17  \\
				B97M-V    & mGGA  & 11.61  & 8.44   & 14.39  & 71.08  & -3.81 &  & MN12-SX       & hmGGA  & 27.11  & 5.26   & 39.26  & 300.82 & 1.92  \\
				M06-L     & mGGA  & 12.09  & 9.17   & 14.73  & 79.4   & -2.42 &  &               &        &        &        &        &        &       \\
				MN15-L    & mGGA  & 12.19  & 10.37  & 13.99  & 54.08  & -2.6  &  & DSD-PBEPBE-D3 & dhGGA  & 3.76   & 2.64   & 4.71   & 20.41  & 0.1   \\
				MN12-L    & mGGA  & 53.24  & 7.59   & 77.47  & 645.34 & 2.93  &  & XYGJ-OS       & dhGGA  & 4.19   & 2.17   & 5.68   & 19.41  & -0.11 \\
				&       &        &        &        &        &       &  & B2GPPLYP      & dhGGA  & 4.36   & 2.38   & 5.84   & 31.58  & 0.54  \\
				SOGGA11-X & hGGA  & 4.84   & 3.78   & 5.82   & 23.93  & 0.69  &  & $\omega$B97X-2       & dhGGA  & 4.52   & 3.31   & 5.6    & 22.37  & 0.87  \\
				$\omega$B97X-V   & hGGA  & 5.07   & 4.32   & 5.8    & 18.82  & 1.81  &  & B2PLYP        & dhGGA  & 5.31   & 2.98   & 7.08   & 36.19  & -0.39 \\
				PBE0      & hGGA  & 5.18   & 4.52   & 5.85   & 22.25  & -0.32 &  &               &        &        &        &        &        &       \\
				B97-2     & hGGA  & 5.34   & 4.5    & 6.16   & 26.92  & -0.58 &  & PWPB95-D3     & dhmGGA & 3.61   & 2.52   & 4.54   & 18.14  & -0.29 \\
				HSEHJS    & hGGA  & 5.35   & 4.56   & 6.14   & 22.64  & -0.45 &  & PTPSS-D3      & dhmGGA & 4.74   & 3.62   & 5.76   & 19.33  & -0.77
			\end{tabular}}
			
			\caption{{RMS {regularized} errors (RMSE), Mean {regularized} errors (ME) and Maximum absolute {regularized} errors (MAX) for selected electronic structure methods over the 152 species at equilibrium, expressed as percentages. The RMSE of the spin-polarized (SP) and non-spin polarized (NSP) subsets of the dataset are also reported separately. The functional with the lowest and highest cumulative RMSE in each class is reported, along with other widely used functionals within the same class. WFT stands for wave function theory, LSDA for local spin density approximation, GGA for generalized gradient approximation, mGGA for meta-GGA, hGGA for hybrid GGA, hmGGA for hybrid meta-GGA, dhGGA for double hybrid GGA and dhmGGA for double hybrid meta-GGA.}}
			\label{my-label}
			\vspace{-10 pt}}
	\end{table*}

The regularized relative error metric defined in the preceding section was used to evaluate the performance of HF, MP2, CCSD and 76 DFAs against the CCSD(T) benchmark. The errors for a selection of popular and new DFAs are given in Table \ref{my-label}, while Table \ref{alldata} contains the benchmark numbers for all 152 species, along with the values predicted by the functional with the lowest cumulative RMSE in each class. The full list of all errors for all methods is provided in the Supporting Information, along with all computed dipole moments.

	{\renewcommand\arraystretch{0.6}
		\setlength\tabcolsep{2pt}	
		\scriptsize{\begin{longtable}[htb!]{lrrrrrr|llrrrrrr}
\caption{{Benchmark CCSD(T)/CBS $\mu$ for all 152 species at equilibrium geometry, along with the values predicted by the functional with lowest cumulative RMSE at each rung of Jacob's ladder. These functionals are SVWN1RPA (Rung 1), N12 (Rung 2), mBEEF (Rung 3), SOGGA11-X (Rung 4) and PWPB95-D3 (Rung 5) respectively.}}
\label{alldata}\\
NSP Species   & CCSD(T) & Rung 1   & Rung 2    & Rung 3   & Rung 4 & Rung 5 &  & SP Species     & CCSD(T) & Rung 1   & Rung 2    & Rung 3   & Rung 4 & Rung 5 \\
\endfirsthead
\multicolumn{15}{c}%
{{\bfseries Table \thetable\ continued from previous page}} \\
NSP Species   & CCSD(T) & Rung 1   & Rung 2    & Rung 3   & Rung 4 & Rung 5 &  & SP Species     & CCSD(T) & Rung 1   & Rung 2    & Rung 3   & Rung 4 & Rung 5\\
\endhead
AlF           & 1.4729  & 1.4752 & 1.4252 & 1.2938 & 1.4011 & 1.4699        &  & AlH$_2$        & 0.4011  & 0.3539  & 0.1894  & 0.4116  & 0.215  & 0.4271        \\
BF            & 0.8194  & 1.0355 & 0.8506 & 1.0184 & 0.9236 & 0.9125        &  & BeH            & 0.2319  & 0.2231  & 0.3462  & 0.3205  & 0.2019 & 0.2845        \\
BH$_2$Cl      & 0.6838  & 0.4333 & 0.5882 & 0.526  & 0.6625 & 0.6071        &  & BH             & 1.4103  & 1.5868  & 1.5077  & 1.3301  & 1.5801 & 1.5314        \\
BH$_2$F       & 0.8269  & 0.6497 & 0.784  & 0.6633 & 0.802  & 0.7754        &  & BH$_2$         & 0.5004  & 0.5317  & 0.5204  & 0.4616  & 0.5777 & 0.51          \\
BHCl$_2$      & 0.6684  & 0.4777 & 0.5898 & 0.552  & 0.6507 & 0.6123        &  & BN             & 2.0366  & 1.8803  & 1.9069  & 2.0663  & 2.1277 & 2.0623        \\
BHF$_2$       & 0.9578  & 0.8136 & 0.9227 & 0.8082 & 0.9349 & 0.9137        &  & BO             & 2.3171  & 2.2235  & 2.4536  & 2.3479  & 2.4548 & 2.3614        \\
CH$_2$BH      & 0.6238  & 0.4916 & 0.5813 & 0.5548 & 0.6984 & 0.6459        &  & BS             & 0.7834  & 0.6292  & 0.8593  & 0.8979  & 0.8797 & 0.7991        \\
CH$_2$BOH     & 2.2558  & 2.2699 & 2.2448 & 2.2801 & 2.3161 & 2.2911        &  & C$_2$H         & 0.7601  & 0.7787  & 0.6924  & 0.7509  & 0.8184 & 0.7798        \\
CH$_3$BH$_2$  & 0.5751  & 0.7712 & 0.6669 & 0.684  & 0.6216 & 0.624         &  & C$_2$H$_3$     & 0.6867  & 0.7805  & 0.6878  & 0.6619  & 0.7236 & 0.7017        \\
CH$_3$BO      & 3.6779  & 3.7118 & 3.7988 & 3.7424 & 3.9061 & 3.7199        &  & C$_2$H$_5$    & 0.314   & 0.4466  & 0.3695  & 0.3105  & 0.3354 & 0.3328        \\
CH$_3$Cl      & 1.8981  & 1.8568 & 1.897  & 1.8662 & 1.907  & 1.8825        &  & CF             & 0.6793  & 0.9517  & 0.717   & 0.8345  & 0.7215 & 0.7345        \\
CH$_3$F       & 1.8083  & 1.6947 & 1.785  & 1.6771 & 1.7613 & 1.7806        &  & CF$_2$         & 0.5402  & 0.7775  & 0.5648  & 0.6791  & 0.57   & 0.5863        \\
CH$_3$Li      & 5.8304  & 5.5556 & 5.3352 & 5.7807 & 5.8242 & 5.7122        &  & CH             & 1.4328  & 1.4732  & 1.4106  & 1.3465  & 1.4561 & 1.4502        \\
CH$_3$NH$_2$  & 1.3876  & 1.3628 & 1.3415 & 1.3053 & 1.3561 & 1.3696        &  & CH$_2$F        & 1.3796  & 1.172   & 1.3094  & 1.2526  & 1.3276 & 1.3379        \\
CH$_3$OH      & 1.7091  & 1.656  & 1.673  & 1.5978 & 1.6682 & 1.6881        &  & CH$_2$NH       & 2.0673  & 2.0032  & 1.9875  & 1.9462  & 2.0464 & 2.0283        \\
CH$_3$SH      & 1.5906  & 1.5968 & 1.609  & 1.5562 & 1.6198 & 1.5806        &  & CH$_2$PH       & 0.8748  & 0.8903  & 0.9728  & 0.9572  & 0.9312 & 0.884         \\
ClCN          & 2.8496  & 3.0204 & 2.9294 & 2.939  & 2.9904 & 2.9123        &  & CH$_2$-singlet & 1.4942  & 1.4737  & 1.3844  & 1.2784  & 1.4729 & 1.7652        \\
ClF           & 0.8802  & 0.7762 & 0.8347 & 0.8115 & 0.8686 & 0.874         &  & CH$_2$-triplet & 0.5862  & 0.6229  & 0.578   & 0.5607  & 0.6406 & 0.6002        \\
CO            & 0.1172  & 0.2226 & 0.0522 & 0.1103 & 0.0254 & 0.1163        &  & CH$_3$O        & 2.0368  & 2.3154  & 2.1345  & 2.0016  & 2.0137 & 2.0463        \\
CS            & 1.9692  & 2.0917 & 1.8248 & 1.8529 & 1.8696 & 1.9787        &  & ClO$_2$        & 1.8627  & 1.7039  & 1.7332  & 1.7186  & 1.8531 & 1.8016        \\
CSO           & 0.7327  & 0.8688 & 0.8706 & 0.7821 & 0.7902 & 0.7857        &  & CN             & 1.4318  & 1.1278  & 1.2966  & 1.3747  & 1.4253 & 1.2342        \\
FCN           & 2.1756  & 2.3212 & 2.1962 & 2.2773 & 2.3025 & 2.2127        &  & FCO            & 0.7678  & 0.7444  & 0.8032  & 0.8317  & 0.8662 & 0.7682        \\
FNO           & 1.6971  & 1.4487 & 1.5165 & 1.5556 & 1.6714 & 1.6628        &  & FH-BH$_2$      & 2.973   & 3.1479  & 3.1095  & 2.9501  & 3.0847 & 3.0079        \\
H$_2$O        & 1.8601  & 1.8587 & 1.8615 & 1.798  & 1.8593 & 1.8584        &  & FH-NH$_2$      & 4.6265  & 4.6766  & 4.6782  & 4.5646  & 4.6246 & 4.6374        \\
H$_2$O-H$_2$O & 2.7303  & 2.8186 & 2.7875 & 2.6932 & 2.7459 & 2.7476        &  & FH-OH          & 3.3808  & 3.2151  & 3.4185  & 3.3352  & 3.3974 & 3.3969        \\
H$_2$O-NH$_3$ & 3.5004  & 3.6318 & 3.5676 & 3.4594 & 3.5063 & 3.5198        &  & H$_2$CN        & 2.4939  & 2.5778  & 2.4593  & 2.4468  & 2.4918 & 2.4896        \\
H$_2$S-H$_2$S & 0.9181  & 1.0983 & 1.0377 & 0.9725 & 0.9652 & 0.94          &  & H$_2$O-Al      & 4.3573  & 4.1889  & 4.1944  & 4.3657  & 4.4005 & 4.3355        \\
H$_2$S-HCl    & 2.1328  & 2.382  & 2.3232 & 2.2473 & 2.2066 & 2.1655        &  & H$_2$O-Cl      & 2.2383  & 3.2691  & 3.128   & 2.8502  & 2.4851 & 2.4656        \\
HBH$_2$BH     & 0.8429  & 0.9077 & 0.7993 & 0.8118 & 0.8334 & 0.8378        &  & H$_2$O-F       & 2.1875  & 3.3613  & 3.0961  & 2.8432  & 2.4333 & 2.4568        \\
HBO           & 2.7322  & 2.63   & 2.7643 & 2.7158 & 2.9037 & 2.7397        &  & H$_2$O-Li      & 3.6184  & 1.8091  & 1.9798  & 3.155   & 3.6389 & 3.181         \\
HBS           & 1.3753  & 1.203  & 1.4219 & 1.4297 & 1.5598 & 1.3769        &  & HCHS           & 1.7588  & 1.6609  & 1.7713  & 1.7892  & 1.8357 & 1.7353        \\
HCCCl         & 0.5009  & 0.2702 & 0.3622 & 0.3546 & 0.4157 & 0.429         &  & HCO            & 1.6912  & 1.6065  & 1.674   & 1.6358  & 1.7711 & 1.6855        \\
HCCF          & 0.7452  & 0.5287 & 0.6683 & 0.5839 & 0.6565 & 0.6969        &  & HCP            & 0.3542  & 0.3388  & 0.485   & 0.4311  & 0.3545 & 0.3854        \\
HCHO          & 2.3927  & 2.2924 & 2.3662 & 2.299  & 2.4636 & 2.3907        &  & HNO            & 1.6536  & 1.5793  & 1.5987  & 1.5115  & 1.666  & 1.6377        \\
HCl           & 1.1055  & 1.1083 & 1.1207 & 1.1071 & 1.1185 & 1.0987        &  & HNO$_2$        & 1.9345  & 1.9524  & 1.9887  & 1.9116  & 1.9706 & 1.9165        \\
HCl-HCl       & 1.7766  & 1.8955 & 1.8682 & 1.8294 & 1.8189 & 1.7869        &  & HNS            & 1.4062  & 1.3789  & 1.3521  & 1.323   & 1.3756 & 1.3814        \\
HCN           & 3.0065  & 3.004  & 2.9891 & 2.9882 & 3.0716 & 3.0179        &  & HO$_2$         & 2.1659  & 2.3119  & 2.198   & 2.098   & 2.1214 & 2.1795        \\
HCNO          & 2.956   & 2.4974 & 2.6608 & 2.6265 & 2.9434 & 2.7715        &  & HOF            & 1.9168  & 1.8721  & 1.9261  & 1.8763  & 1.9261 & 1.9056        \\
HCOF          & 2.1169  & 2.0352 & 2.0962 & 2.0478 & 2.1781 & 2.1141        &  & HPO            & 2.6291  & 2.3349  & 2.3454  & 2.3221  & 2.5614 & 2.4261        \\
HCONH$_2$     & 3.9152  & 3.9154 & 3.9307 & 3.8721 & 4.0033 & 3.9369        &  & LiN            & 7.0558  & 6.6258  & 5.5154  & 6.8321  & 7.014  & 6.969         \\
HCOOH         & 1.3835  & 1.4209 & 1.4402 & 1.4712 & 1.4992 & 1.4237        &  & N$_2$H$_2$     & 2.8771  & 2.8749  & 2.8102  & 2.7213  & 2.8434 & 2.8593        \\
HF            & 1.8059  & 1.7973 & 1.8213 & 1.7621 & 1.8109 & 1.806         &  & NaLi           & 0.4837  & -0.1785 & 0.456   & 0.2608  & 0.2444 & 0.4564        \\
HF-HF         & 3.3991  & 3.4486 & 3.4643 & 3.3529 & 3.4228 & 3.4135        &  & NCl            & 1.1279  & 1.3884  & 1.0582  & 1.0968  & 1.0759 & 1.1342        \\
HN$_3$        & 1.6603  & 1.822  & 1.7601 & 1.6822 & 1.6441 & 1.6828        &  & NCO            & 0.7935  & 0.8606  & 0.7605  & 0.8335  & 0.7554 & 0.7925        \\
HNC           & 3.0818  & 3.1302 & 2.9563 & 2.9645 & 2.9632 & 3.0497        &  & NF             & 0.0671  & 0.3429  & 0.0529  & 0.1487  & 0.058  & 0.0636        \\
HNCO          & 2.0639  & 2.0372 & 2.0656 & 2.0007 & 2.0634 & 2.0465        &  & NF$_2$         & 0.1904  & -0.0079 & 0.198   & 0.1429  & 0.2338 & 0.1947        \\
HOCl          & 1.5216  & 1.531  & 1.5464 & 1.4999 & 1.5364 & 1.5223        &  & NH             & 1.5433  & 1.5321  & 1.4952  & 1.4847  & 1.5172 & 1.5267        \\
HOCN          & 3.7998  & 3.9528 & 3.8754 & 3.885  & 3.9417 & 3.8466        &  & NH$_2$         & 1.7853  & 1.7855  & 1.7505  & 1.7188  & 1.7669 & 1.7749        \\
HOOH          & 1.5732  & 1.5702 & 1.5796 & 1.5294 & 1.5741 & 1.5738        &  & NO             & 0.1271  & 0.2427  & 0.0952  & 0.1105  & 0.0638 & 0.146         \\
LiBH$_4$      & 6.1281  & 5.9182 & 5.9522 & 6.0424 & 6.0772 & 6.0561        &  & NO$_2$         & 0.335   & 0.2782  & 0.3299  & 0.2812  & 0.3891 & 0.3205        \\
LiCl          & 7.096   & 6.8398 & 6.9066 & 6.9944 & 7.076  & 7.0191        &  & NOCl           & 2.0773  & 1.6196  & 1.611   & 1.8626  & 2.0024 & 1.964         \\
LiCN          & 6.9851  & 6.8143 & 6.8206 & 6.911  & 6.9746 & 6.9434        &  & NP             & 2.8713  & 2.8217  & 2.6854  & 2.7355  & 2.9444 & 2.8042        \\
LiF           & 6.2879  & 6.0928 & 6.1585 & 6.1847 & 6.2636 & 6.2377        &  & NS             & 1.8237  & 1.87    & 1.6785  & 1.6518  & 1.7763 & 1.8411        \\
LiH           & 5.8286  & 5.6334 & 5.4829 & 5.8214 & 5.8515 & 5.7695        &  & O$_3$          & 0.5666  & 0.6309  & 0.5996  & 0.5046  & 0.4552 & 0.601         \\
LiOH          & 4.5664  & 4.3294 & 4.3895 & 4.4928 & 4.5533 & 4.5033        &  & OCl            & 1.279   & 1.4681  & 1.2892  & 1.1782  & 1.1946 & 1.3296        \\
N$_2$H$_4$    & 2.7179  & 2.729  & 2.6778 & 2.6137 & 2.6907 & 2.705         &  & OF             & 0.0205  & -0.2508 & -0.0133 & -0.0263 & 0.0912 & 0.0357        \\
NaCl          & 9.0066  & 8.5165 & 8.4581 & 8.7699 & 8.9231 & 8.8444        &  & OF$_2$         & 0.3252  & 0.2662  & 0.3414  & 0.3087  & 0.357  & 0.3325        \\
NaCN          & 8.8903  & 8.5782 & 8.5639 & 8.741  & 8.8518 & 8.8076        &  & OH             & 1.655   & 1.6503  & 1.6479  & 1.6053  & 1.6503 & 1.6522        \\
NaF           & 8.1339  & 7.7791 & 7.7797 & 7.9124 & 8.0702 & 8.0393        &  & PCl            & 0.5657  & 0.2862  & 0.3479  & 0.4656  & 0.522  & 0.5253        \\
NaH           & 6.3966  & 5.7148 & 5.569  & 6.1613 & 6.634  & 6.2937        &  & PF             & 0.8104  & 0.6164  & 0.7254  & 0.6623  & 0.7757 & 0.7873        \\
NaOH          & 6.769   & 6.5013 & 6.5337 & 6.677  & 6.7397 & 6.6964        &  & PH             & 0.4375  & 0.4447  & 0.4508  & 0.45    & 0.4563 & 0.4205        \\
NH$_2$Cl      & 1.9468  & 1.9474 & 1.9821 & 1.959  & 1.9794 & 1.9474        &  & PH$_2$         & 0.5472  & 0.5763  & 0.5785  & 0.539   & 0.5731 & 0.5275        \\
NH$_2$F       & 2.2688  & 2.2126 & 2.2753 & 2.2186 & 2.2693 & 2.2681        &  & PO             & 1.9617  & 1.8382  & 1.8503  & 1.8966  & 2.0904 & 1.961         \\
NH$_2$OH      & 0.7044  & 0.6583 & 0.6966 & 0.6768 & 0.6937 & 0.7013        &  & PO$_2$         & 1.4426  & 1.2894  & 1.3182  & 1.366   & 1.4784 & 1.3926        \\
NH$_3$        & 1.5289  & 1.5291 & 1.5056 & 1.4786 & 1.5191 & 1.5235        &  & PPO            & 1.8812  & 1.8136  & 1.7997  & 1.7817  & 1.9327 & 1.9082        \\
NH$_3$-BH$_3$ & 5.281   & 5.276  & 5.2421 & 5.2833 & 5.2871 & 5.2772        &  & PS             & 0.6825  & 0.4854  & 0.5238  & 0.6702  & 0.7253 & 0.5932        \\
NH$_3$-NH$_3$ & 2.1345  & 2.2246 & 2.167  & 2.1089 & 2.1412 & 2.1465        &  & SCl            & 0.069   & -0.2227 & -0.1396 & 0.0026  & 0.048  & 0.0115        \\
NH$_3$O       & 5.3942  & 5.161  & 5.1227 & 5.0837 & 5.3136 & 5.3235        &  & SF             & 0.8139  & 0.5787  & 0.7018  & 0.6704  & 0.7955 & 0.784         \\
OCl$_2$       & 0.5625  & 0.5223 & 0.515  & 0.5075 & 0.5487 & 0.562         &  & SH             & 0.7727  & 0.7809  & 0.7883  & 0.7712  & 0.786  & 0.7636        \\
P$_2$H$_4$    & 0.9979  & 1.0139 & 1.0209 & 0.9576 & 1.0353 & 0.9651        &  & SiH            & 0.1138  & 0.1206  & 0.1495  & 0.1406  & 0.1658 & 0.0953        \\
PH$_2$OH      & 0.6836  & 0.6556 & 0.661  & 0.6719 & 0.6551 & 0.6993        &  & SO-triplet     & 1.5606  & 1.4131  & 1.3901  & 1.4225  & 1.5576 & 1.5225        \\
PH$_3$        & 0.6069  & 0.6484 & 0.6494 & 0.5871 & 0.6309 & 0.5839        &  &                &         &         &         &         &        &               \\
PH$_3$O       & 3.7704  & 3.5945 & 3.623  & 3.5384 & 3.797  & 3.7092        &  &                &         &         &         &         &        &               \\
S$_2$H$_2$    & 1.1425  & 1.1392 & 1.1415 & 1.1053 & 1.1561 & 1.1202        &  &                &         &         &         &         &        &               \\
SCl$_2$       & 0.3891  & 0.2986 & 0.334  & 0.3631 & 0.3824 & 0.382         &  &                &         &         &         &         &        &               \\
SF$_2$        & 1.0555  & 0.8581 & 0.9825 & 0.8846 & 1.0232 & 1.0318        &  &                &         &         &         &         &        &               \\
SH$_2$        & 0.9939  & 1.0121 & 1.0186 & 0.9756 & 1.0136 & 0.9822        &  &                &         &         &         &         &        &               \\
SiH$_3$Cl     & 1.3645  & 1.2687 & 1.2974 & 1.2286 & 1.3336 & 1.3094        &  &                &         &         &         &         &        &               \\
SiH$_3$F      & 1.3123  & 1.2486 & 1.3284 & 1.1861 & 1.3013 & 1.2835        &  &                &         &         &         &         &        &               \\
SiO           & 3.1123  & 2.9985 & 2.9847 & 2.9984 & 3.2812 & 3.1155        &  &                &         &         &         &         &        &               \\
SO$_2$        & 1.6286  & 1.5268 & 1.5616 & 1.5477 & 1.6839 & 1.6246        &  &                &         &         &         &         &        &              
\end{longtable}}}

\subsection{Performance of Wave function Theory}
As expected, CCSD is the most effective wave function theory (WFT) in reproducing the CCSD(T) benchmark, giving a regularized RMSE of 3.95\% over the whole dataset. This difference is large enough, however, to clearly show that CCSD itself is not a suitable replacement, and, by implication, that CCSD densities themselves have limitations. Furthermore, there exist cases like the PS radical where CCSD deviates from CCSD(T) by as much as 13\%, raising some questions about whether CCSD(T) itself is adequate as a benchmark for those species (as it only adds a perturbative triples correction to CCSD). However, comparison between CCSD(T) and the more robust (and computationally expensive) CCSD(2)\cite{gwaltney2001second} method at the aug-cc-pCVTZ level for species where the CCSD/CCSD(T) regularized deviation is $>6\%$ reveals that CCSD(2) and CCSD(T) are generally within 1\% of each other (full table given in Supporting Information), indicating that the latter is adequate for our purposes. Furthermore, comparison to available experimental values for non-hydrogen containing species (where vibrational averaging is expected to have the least impact) indicates no glaring issues with the CCSD(T) benchmark. These two consistency checks thus collectively suggest that CCSD(T) is likely to be accurate to $\approx 1\%$ even in the worst of cases, and is probably much more accurate for better behaved species, allowing it to serve as a benchmark for our purposes. 

The surprisingly poor performance of MP2 is a consequence of N-representability violation in unrestricted MP2\cite{Kurlancheek2009}, which leads to some very poor numbers for some even-electron singlet systems (like HPO) where spin symmetry is broken, as well as for radicals like ClO$_2$. The species in the spin-unpolarized NSP subset do not cause much trouble for MP2, and the resulting regularized RMSE of $3.96\%$ is very close to the CCSD RMSE of $3.01\%$ over this subset. On the other hand, the massive $54.63\%$ RMSE over the more challenging spin-polarized SP subset indicates that unrestricted MP2 is a poor choice for estimating $\mu$ for such species. Restricted MP2 (RMP2) however performs quite well, giving a much better RMSE of of $9.84\%$ over the SP subset, despite operating on a restricted reference that is higher in energy than the symmetry broken solution. RMP2 consequently has an overall RMS error of 7.3\% over the whole dataset, which, as we shall see, is not too bad relative to the best {hybrid} DFAs. 

Not unexpectedly, HF theory performs least well out of all the WFT methods, giving a regularized RMS error of 18.61\%. Using purely restricted HF does not however lead to a significant deterioration in quality, as it only adds $\approx1.5$\% to the regularized RMS error. It is well known that neglect of correlation in HF theory typically leads to systematic overestimation of dipole moments. Qualitatively, this can be viewed as arising from the fact that antibonding orbitals, which are partly occupied due to correlation, but are empty at the HF level, have polarization opposite to bonding orbitals.

\subsection{Performance of DFT}
There are a number of general features that emerge from inspection of Table \ref{my-label} that are likely to be transferable beyond our specific dataset. First, it is striking that the total RMSE for the best DFA in each class decreases as one ascends Jacob's ladder. This is a nice validation of the fact that there is additional physical content at each level of the ladder. If that additional physical content is used effectively, it can contribute to higher levels of accuracy in the dipole moments, and by inference, the electron density. By contrast, it is also striking that the worst GGA, mGGA and hybrid functionals actually have larger cumulative RMSE than the LSDA. From this we can infer that the additional physical content contained at each higher level of the ladder also leads to greater flexibility in the design of the functional, which, indeed, can degrade the accuracy of the dipole moments relative to LSDA which is a simple parameter-free model. {This is consistent with the general conclusions in Refs [\onlinecite{medvedev2017density,brorsen2017accuracy}], which also found that the best performing hybrid functionals yielded superior density predictions than the best local functionals, while several modern functionals (both hybrid and local) gave performance comparable to or worse than LSDA}. The overall relative performance of functionals in predicting dipole moments was thus not too dissimilar to the performance in predicting density\cite{medvedev2017density,brorsen2017accuracy}, despite the differences in methodology.

{The best functionals for dipole moment prediction are double hybrids, led by PWPB95-D3\cite{goerigk2010efficient} (3.61\% regularized RMSE)
and DSD-PBEPBE-D3\cite{kozuch2013spin} not too far behind (3.76\% regularized RMSE). B2PLYP is the only double hybrid functional tested to have a regularized RMSE in excess of 5\%, and seven out of the remaining eight yield regularized RMSE around 4.5\% or less. This indicates that most double hybrid functionals tested have performance comparable to CCSD for predicting dipole moments, and the top two appear to be better overall for our dataset. This high level of accuracy is somewhat surprising as all the double hybrid calculations at equilibrium geometry were done using restricted orbitals, and thus poor performance appeared nearly inevitable for spin-polarized systems. The errors over the SP subset are indeed substantially larger than the errors over the NSP subset (being nearly twice as large for quite a few functionals), but are still either less than or comparable to corresponding errors for hybrid functionals using unrestricted orbitals (and are significantly less than the corresponding RMP2 errors), resulting in better performance overall. Many of the double hybrids also have very low RMSE over the NSP subset (2-3\%, often substantially better than CCSD), which helps drive down the cumulative error significantly as well. Overall, double hybrid functionals give across the board good performance despite being constrained to use restricted orbitals, and should always be employed to estimate dipole moments of spin unpolarized systems where computationally feasible. They are also very likely to give quite good predictions for spin polarized systems when restricted orbitals are employed to prevent N-representability violations, though there exist a few cases like singlet CH$_2$ where this approach does not yield satisfactory results.}  

Hybrid functionals also perform quite well in predicting dipole moments relative to the benchmark, with SOGGA11-X\cite{sogga11x} being the best performer (4.84\% RMSE), and no fewer than 22 functionals with less than 6\% RMSE. Such functionals are thus quite competitive with CCSD (3.95\% RMSE) and double hybrids. This close relative performance of many functionals at hybrid and {double hybrid} levels  indicates that good performance for dipole moments is possible from many functionals. It is interesting that the best hybrid meta-GGAs (hmGGAs) do not substantially outperform the best hybrid GGAs (hGGAs). This suggests that the additional flexibility in current hmGGAs has not necessarily resulted in substantial improvements in densities, which is perhaps not all that surprising in light of the vastness of the meta-GGA functional space\cite{b97mv}. Recently developed {hybrid meta-GGAs} such as $\omega$B97M-V and SCAN0 are in the top cohort, though they are not statistically the best. Furthermore, most of the best-performing hybrids exhibit RMSEs that are between 30\% and 100\% larger for the SP cases than the NSP cases. Indeed apart from the top two, the next 10 best-performing functionals would reorder considerably if one considered only the NSP cases.


The best local functionals, mBEEF\cite{wellendorff2014mbeef} and SCAN\cite{SCAN}, are significantly less accurate than the best hybrids, with RMS errors of 7.56\% and 8.44\% over the whole database. Local functionals typically underestimate dipole moment magnitudes, while HF and functionals with 100\% exact exchange tend to overestimate. It is interesting that the best-performing local functionals are all quite recently developed functionals -- the N12 GGA, and the mBEEF, SCAN and MS2 mGGAs. Other recently developed semi-empirical functionals, such as B97M-V and MN15-L perform more poorly for dipole moments, so this provides some evidence that fitting to relative energies can lead to loss of accuracy in other properties. For the NSP subset, the best GGA (N12) outperforms the best mGGA (mBEEF), but this result reverses for the more challenging SP subset, as well as the total statistics.  As in hybrids, the gap in performance between the SP and NSP subsets for the best functionals is generally smaller than for the wave function-based MP2 method, and is more comparable to CCSD. The statistics for the SP subset (both for locals and hybrids) are quite heavily influenced by outliers, such as those discussed below.

{A list of the 11 most challenging species overall is given in Table \ref{difficult-species}. Somewhat unsurprisingly, 8 of the 11 are from the SP subset, which is a consequence of the best functionals being double hybrids that we have constrained to operate on restricted references alone}. Perhaps the most challenging species is NaLi. WFT methods all agree that the polarity is Na$^+$Li$^{-}$ and the CCSD(T) $\mu=0.48$ D is close to the experimental value of $0.46$ D\cite{graff1972electric}. Many functionals (both locals like BLYP\cite{b88,lyp} and B97M-V\cite{b97mv} and hybrids like MN15\cite{MN15} and BMK\cite{bmk}) however predict the opposite polarity without any apparent systematic reason. Furthermore, BMK\cite{bmk}, MN12-L\cite{mn12l}, MN12-SX\cite{mn12l}, M11\cite{m11}, M06-HF\cite{m06hf} and SOGGA11\cite{sogga11} predict errors in excess of 100\%, with BMK\cite{bmk} dropping from being one of the very best (4.41\% RMS error otherwise) to one of the worst due to this one data-point, and MN12-L\cite{mn12l} predicting an absurd dipole moment of $6.94$ D! Fortunately, some functionals do well: $\omega$B97X-V\cite{wb97xv}, $\omega$B97X-rV\cite{wb97xv,b97mrv}, $\omega$B97X-D3\cite{wB97XD3} and PWB6K\cite{pw6b95} predict errors around $1\%$ or less.

\begin{table}[]
	\centering
	\begin{tabular}{l|l}
		\textbf{NSP}  & \textbf{SP}  \\
		\hline
		BF& H$_2$O-F \\
		HCCF & NaLi\\
		HCCCl & H$_2$O-Cl\\
		 & H$_2$O-Li\\
		 & CF\\
		 & BS\\
		&CF$_2$\\
		&NOCl\\
	\end{tabular}
	\caption{The 11 most difficult species in the dataset. These were selected on the basis of the first quartile of absolute DFA errors for each species, with the selected species giving $\ge 5.5\%$ first quartile error. This corresponded to a break in the distribution. Coincidentally, this is also close to the first quartile of RMSEs for functionals considered ($5.64\%$).}
	\label{difficult-species}
\end{table}


{The three challenging NSP species are BF and triple bond containing HCCF and HCCCl. There are some striking systematics in functional performance in these challenging spin unpolarized cases. 86 out of 88 functionals overestimate $\mu ({\text{BF}})$, 82 of 88 underestimate $\mu ({\text{HCCF}})$ and 84 of 88 underestimate $\mu ({\text{HCCCl}})$. These systematics suggest that a hitherto unrecognized delocalization error may be at play here, potentially connected to the C$\equiv$C$-$X moiety. Other spin unpolarized boron containing species like CH$_3$BH$_2$ and BH$_2$F are problematic for functionals from the first four rungs of Jacob's ladder (having first quartile errors $>5.5\%$ when double hybrids are excluded), which also hint at potential delocalization errors involving bonds to boron.} Surprisingly, the strongly correlated O$_3$ molecule gives relatively little trouble, although most functionals are forced to break spin-symmetry.


The performance for open-shell and incomplete valence species is generally not significantly poorer than for closed shell species for most {functionals from Rungs 1-4 of Jacob's ladder}, although some radical species are challenging for all functionals, notably the complexes of Li, F and Cl with H$_2$O (the equivalent Al complex however is surprisingly well behaved), and some small fluorine-containing radicals, such as CF$_2$ and CF. HF theory predicts incorrect polarity for some species like CO, NF and NO, causing functionals with large amounts of exact exchange (like M06-HF\cite{m06hf} or {50\% exact exchange containing} PBE50) to also reverse the dipole direction. The opposite problem occurs for some cases like OF, where HF theory gets the direction right but overestimates the magnitude significantly, while most local functionals like PBE\cite{PBE} and BLYP\cite{b88,lyp} incorrectly obtain the reverse polarity.

\subsection{Basis Set Convergence}
	HF dipole moments converge quite rapidly with increasing basis set size, with the regularized RMS deviation between aug-cc-pCVQZ and aug-cc-pCV5Z values being 0.04$\%$ (with a maximum regularized deviation of $0.19\%$). Even aug-cc-pCVTZ $\mu$ have a regularized RMS deviation of only $0.4\%$ relative to aug-cc-pCV5Z numbers (with a maximum regularized deviation of $1\%$) indicating that the aug-cc-pCV5Z values are essentially at the CBS limit. 
	
	Most of the DFAs {from Rungs 1-4 of Jacob's ladder} exhibit similarly rapid convergence, with a regularized RMS difference of $\le 0.1\%$ between aug-pc-4 and aug-pc-3, and $\approx 1\%$ or so between aug-pc-4 and aug-pc-2 (a full list of regularized RMS differences is provided in the Supporting Information). Expensive aug-pc-4 calculations of dipole moments are thus unnecessary for practical purposes, as aug-pc-3 and even aug-pc-2 numbers are likely to be sufficiently close for most functionals (especially considering that {these functionals} typically deviate from benchmark by $>5-6\%$).

\begin{table}[]
	\centering
	\caption{{Percentage regularized differences between aug-pc-4 basis $\mu$ and aug-pc-3/aug-pc-2 $\mu$ for DFT Functionals with $>1\%$ regularized RMS difference between aug-pc-4 and aug-pc-3 values. A full table with deviations for all functionals is supplied in the Supporting Information.}}
	\label{funcs}
	\setlength\tabcolsep{3pt}
	\scriptsize{\begin{tabular}{lrrrrrr}
			Functional & \multicolumn{2}{c}{RMSE} & \multicolumn{2}{c}{ME} & \multicolumn{2}{c}{MAX} \\
			& aug-pc-3    & aug-pc-2   & aug-pc-3   & aug-pc-2  & aug-pc-3   & aug-pc-2   \\
			M08-HX     & 1.06        & 2.29       & 0.04       & 0.5       & 10.54      & 10.13      \\
			BMK        & 1.09        & 1.79       & 0.02       & 0.17      & 9.62       & 17.72      \\
			M06        & 1.3         & 3.6        & 0.01       & -0.3      & 12.77      & 40.75      \\
			M11        & 1.39        & 2.4        & 0.03       & 0.73      & 7.54       & 13.1       \\
			M06-HF     & 2.47        & 4.12       & -0.68      & 0.21      & 13.07      & 39.5       \\
			MN15-L     & 2.68        & 1.61       & 0.06       & 0.26      & 29.63      & 14.02      \\
			M06-L      & 2.93        & 4.58       & -0.31      & -0.95     & 27.82      & 48.77      \\
			M11-L      & 3.94        & 8.32       & -0.19      & 0.06      & 34.18      & 83.57      \\
			MN12-SX    & 5.4         & 9.08       & -0.3       & -0.6      & 62.96      & 78.07      \\
			SOGGA11    & 11.51       & 13.55      & 1.97       & 2.4       & 37.89      & 40.1       \\
			MN12-L     & 18.39       & 7.86       & 1.21       & 0.13      & 223.74     & 91.5      
		\end{tabular}}
	\end{table}
	
	A few functionals {from Rungs 1-4} are however exceptional in having a $1\%$ or larger difference between aug-pc-3 and aug-pc-4 values, suggesting that their density has potentially not yet fully converged even at the quintuple $\zeta$ aug-pc-4 level. A full list of these potentially problematic functionals is given in Table \ref{funcs}, along with the mean and maximum deviations by magnitude. The poor basis set convergence of intermolecular interaction energies for several of the listed functionals has been noted in the past\cite{mardirossian2013characterizing}. It highlights the need to exercise caution when these functionals are being used to estimate molecular properties, due to the lack of an easily approachable convergence limit. SOGGA11 in particular has an inaccessible CBS limit, with 99 species out of 152 having a regularized difference of $>5\%$ between aug-pc-3 and aug-pc-4 $\mu$, resulting in a very high 11\% regularized RMS difference between the two basis sets that is not outlier driven (unlike, say the case of MN12-L---where only 6 species have a $>5\%$ deviation). It thus appears that the electronic density for SOGGA has likely not converged sufficiently even when the very large aug-pc-4 basis is employed.

{Double hybrid functionals exhibit relatively slower basis set convergence compared to other DFAs on account of their dependence on virtual orbitals. The SCF component of $\mu$ for all the tested double hybrids however converge at a rate similar to HF and most other functionals ($0.05-0.25\%$ regularized RMS deviation between aug-cc-pCV5Z and aug-cc-pCVQZ, and $<0.5\%$ regularized RMS difference between aug-cc-pCV5Z and aug-cc-pCVTZ), and aug-cc-pCV5Z numbers are thus expected to be very close to the CBS limit. The $3.5-5\%$ RMS error associated with $\mu$ estimated from double hybrid functionals suggest that the non MP2 component of aug-cc-pCVQZ results are probably adequately close to be CBS for practical purposes. The MP2 component has slower basis set convergence, but the behavior can be well approximated by the extrapolation formula $\mu^{\mbox{corr}}_n=\mu^{\mbox{corr}}_{\infty}+A/n^3$\cite{halkier1999basis}. We therefore recommend that the CBS MP2 component be estimated via extrapolation from aug-cc-pCVQZ and aug-cc-pCVTZ numbers, and be combined with the SCF component at aug-cc-pCVQZ level to obtain a practically useful estimate for $\mu$ that does not require any expensive aug-cc-pCV5Z calculations.}

\section{Dipole moments at non-equilibrium configurations}
\begin{figure}[htb!]
	\centering
	\includegraphics[width=3.25 in]{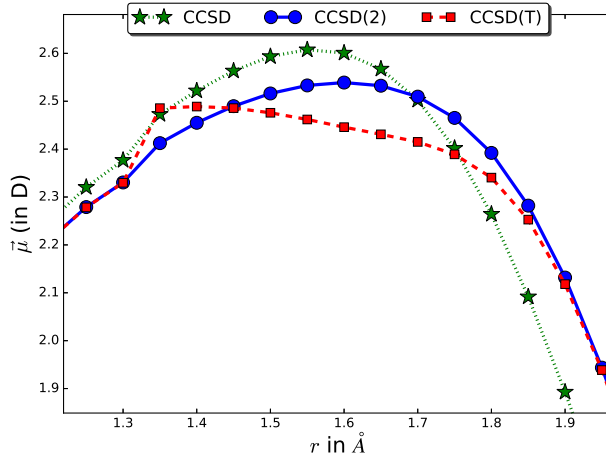}
	\vspace*{-15 pt}
	\caption{Performance of CC methods in predicting $\vec{\mu}$ near Coulson-Fischer point ($r_e=1.35$ \AA) of FH. Note the pronounced kink in the CCSD(T) curve.}
	\label{fig:wft-hf}
\end{figure}
Exploration of non-equilibrium behavior by examining $\vec{\mu}$ along the PES of FH and FCl reveals serious qualitative issues with many of the methods discussed so far.
Just as the energy for bond-stretched configurations is a challenge for electronic structure theory, so too is the dipole moment, because such configurations reflect the effect of stronger electron correlations on the property of interest. Indeed, the accuracy of CCSD(T) itself becomes questionable close to the Coulson-Fischer\cite{coulson1949xxxiv} point of the FH PES due to the appearance of kinks\cite{Kurlancheek2009}. This led us to investigate CCSD(2)\cite{gwaltney2001second} as an alternative benchmark in such situations, due to its renormalization of the one-body terms.\cite{gwaltney2000} The performance of CCSD, CCSD(T) and CCSD(2) around the Coulson-Fischer point of FH dissociation (1.35 {\AA} internuclear separation) is given in Fig \ref{fig:wft-hf}. CCSD and CCSD(2) give reasonable behavior, but CCSD(T) has a kink at the spin polarization transition, indicating that CCSD(2) is more reliable. The two methods however agree at shorter and longer distances, and no other unreasonable behavior is observed. No similar issues arise in the case of the FCl molecule, where CCSD(T) and CCSD(2) yield dipole moment curves that do not demonstrate any unphysical behavior and are in perfect visual agreement. Thus CCSD(T) $\vec{\mu}$ for FCl dissociation can be treated as a benchmark.

\begin{figure}[htb!]
	\centering
	\includegraphics[width=3.25 in]{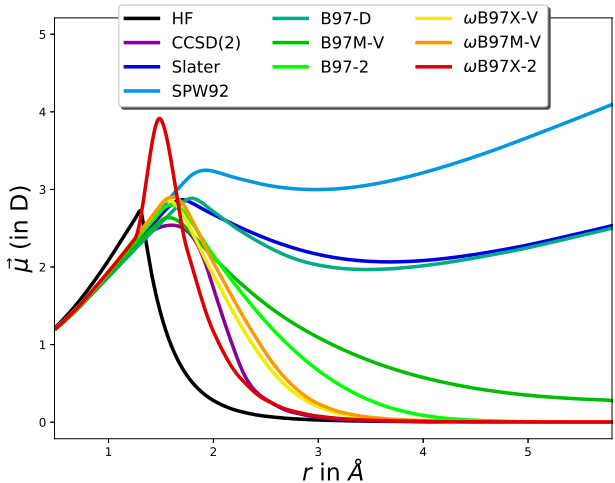}
	\vspace*{-10 pt}
	\caption{Performance of selected DFAs in predicting $\vec{\mu}$ for FH dissociation. $r$ is the internuclear separation.}
	\label{fig:dft-hf}
\end{figure}

The performance of DFAs in predicting $\vec{\mu}$ along the FH PES varies greatly with internuclear separation, $r$, as shown in Fig. \ref{fig:dft-hf}.
The behavior at $r \le r_e$ is qualitatively fine for all methods, but the performance around the Coulson-Fischer point and larger $r$ is much poorer. Local functionals often spuriously predict a residual charge on the atoms even at very large $r$\cite{Ruzsinszky2006,Dutoi2006}, causing $\mu$ to incorrectly grow asymptotically with $r$.
Surprisingly, SPW92/SVWN\cite{PW92,VWN} leaves behind a larger residual atomic charge than pure Slater\cite{Slater} exchange without correlation, as can be seen from relative slopes in Fig \ref{fig:dft-hf}. 
GGAs in general are able to partially reverse this, returning to pure Slater like behavior.

Moving higher up Jacob's ladder to mGGAs alleviates this issue somewhat further, though there are distinct differences between functionals. TPSS\cite{tpss}, revTPSS\cite{revTPSS} and $\tau$-HCTH\cite{thctch} act similarly to GGAs, but SCAN\cite{SCAN}, MS2\cite{ms2}, mBEEF\cite{wellendorff2014mbeef}, B97M-V\cite{b97mv} and B97M-rV\cite{b97mrv} have a smaller residual atomic charge by nearly an order of magnitude (this is not apparent in Fig \ref{fig:dft-hf} as the B97M-V asymptotic regime is not yet reached). Rather surprisingly, MS1\cite{ms2}, MVS\cite{mvs}, revM06-L\cite{revm06l}, M06-L\cite{m06l}, M11-L\cite{m11l}, MN12-L\cite{mn12l} and MN15-L\cite{mn15l} appear to leave behind no partial charge at all.

Hybrid functionals on the other hand, are universally able to eliminate this self-interaction issue, allowing $\mu$ to go to the correct asymptotic limit of zero. The overall decay rate however is typically slower than that of the CCSD(2) benchmark (except for functionals with 100\% exact exchange like HFLYP\cite{sherrill1999performance}, which tend to decay \textit{too} fast like HF), indicating that the asymptotic tail of the $\mu$ curve is problematic for modern DFAs.
Hybrids also tend to overestimate the maximum $\mu$ considerably (as can be seen in Figure \ref{fig:dft-hf}).
These difficulties relate to strong electron correlation effects, and represent challenges for future functional development. Nonetheless, it is encouraging the functionals that are most accurate for chemistry, such as $\omega$B97M-V\cite{wB97MV}, are amongst the most accurate relative to the CCSD(2) benchmark. 

{Double hybrid functionals however do not fare as well in the stretched bond regime. These functionals inherit N-representability violations present in unrestricted MP2, and consequently may exhibit an unphysical spike in the value of $\vec{\mu}$ around the Coulson-Fischer point (as can be seen in Fig \ref{fig:dft-hf}). Using restricted orbitals (as in the equilibrium case) however yielded asymptotically divergent $\vec{\mu}$ in the bond dissociation limit, as these orbitals artificially enhance the ionic character in stretched bonds. Consequently, $\vec{\mu}$ calculated from unrestricted orbitals is the better (albeit flawed) choice for highly stretched bonds, as it correctly decays to $0$ as $r\to\infty$. The decay rate however appears to be too large relative to the reference values, which is likely a consequence of the larger amount of exact exchange typically employed by double hybrids.} 

\begin{figure}[htb!]
	\centering
	\vspace*{-10 pt}
	\includegraphics[width=3.25 in]{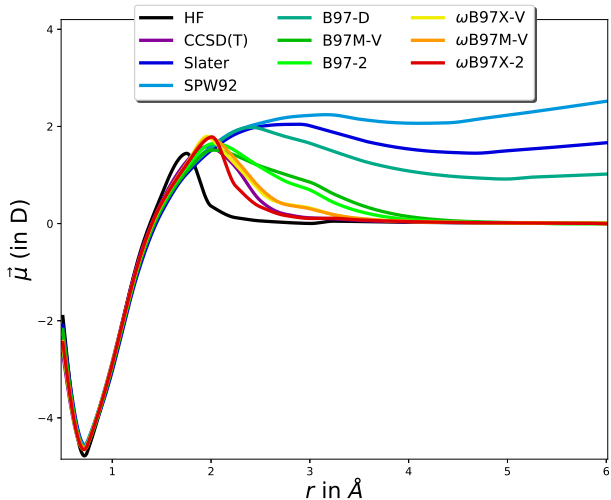}
	\vspace*{-10 pt}
	\caption{Performance of selected DFAs in predicting $\vec{\mu}$ for FCl dissociation. $r$ is the internuclear separation.}
\vspace*{-10 pt}
	\label{fig:dft-clf}
\end{figure}

As shown in Fig. \ref{fig:dft-clf}, the FCl dissociation is less problematic for functionals, despite the larger median DFA error of 6.65\% at equilibrium. LDAs and GGAs still predict a residual charge on the atoms, but the magnitude is significantly lower due to the smaller electronegativity difference between F and Cl (versus F and H). On the other hand, all mGGAs except the aforementioned problematic cases of TPSS\cite{tpss}, revTPSS\cite{tpssh} and $\tau$-HCTH\cite{thctch} (which give GGA like behavior) are able to correctly predict neutral atoms at dissociation, showing that the extra physics in the mGGAs is in principle capable of eliminating a weak manifestation of the self-interaction error. In fact, mGGAs like B97M-V\cite{b97mv} match the performance of hybrids like B97-2 in this case, although the best performers are still hybrids like $\omega$B97X-V\cite{wb97xv}. {None of the double hybrids here displayed a sharp spike in $\mu$, indicating that N-representability was likely not violated in this case despite use of unrestricted orbitals}.

\begin{figure}[htb!]
	\centering
	\vspace*{-10 pt}
	\includegraphics[width=3.25 in]{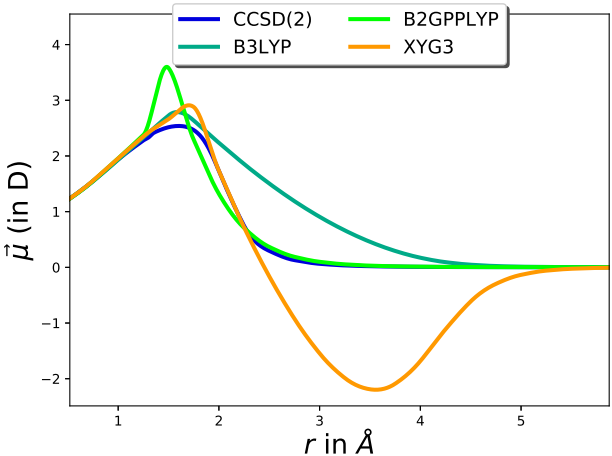}
	\vspace*{-10 pt}
	\caption{{$\vec{\mu}$ predicted by the XYG3 functional for FH dissociation, against the CCSD(2) benchmark and the related B3LYP and B2GPPLYP functionals.}}
	\vspace*{-10 pt}
	\label{fig:xyg3}
\end{figure}

{One interesting phenomenon unique to the XYG3\cite{zhang2009doubly} and XYGJ-OS\cite{zhang2011fast} double hybrid functionals was observed for both the FH and FCl dissociation curves. These two functionals spuriously reversed the polarity of $\vec{\mu}$ by a considerable amount at moderately large $r$, as demonstrated in Fig. \ref{fig:xyg3} for the case of FH dissociation with the XYG3 functional. We are currently investigating the origins of this unphysical behavior and any potential implications it may have for development of double hybrid functionals.} 

\section{Conclusion}
{In summary, we find that density functionals are quite good at predicting ${\mu}$, with the best DFAs having an RMS error of $3.5$-$4.5\%$, which is not particularly different from the CCSD RMS error of $3.95\%$, at a significantly lower computational cost. Most of this success is on account of excellent performance by double hybrid functionals, which yield performance similar to or even slightly superior than CCSD for most species. Hybrid functionals also perform quite well, with 22 functionals having RMS error $<6\%$, which is only about 50\% worse than CCSD at an even lower computational cost}. This success story however hides the fact that many of the best performing hybrid functionals like PBE0 and B97-2 are quite old, and newer functionals like SOGGA11-X\cite{sogga11x} and $\omega$B97X-V\cite{wb97xv} are only comparable in accuracy. Furthermore, hybrid mGGAs give no improvement in  performance compared to the best hybrid GGAs, despite their more sophisticated form. 
This may reflect the vastness of the mGGA functional space and the previous emphasis on improving energetics (where the best hybrid mGGAs are superior to the best hybrid GGAs).

Assuming that errors in $\vec{\mu}$ are an acceptable proxy for errors in electron density, we conclude that {aside from the relatively recent class of double hybrid functionals,} there has not been a great deal of progress in improving electron densities in recent years, which is consistent with other recent studies\cite{brorsen2017accuracy,medvedev2017density}. {While the best performing GGA (N12) and mGGA (mBEEF) are relatively recent and are considerably superior to classics like PBE or TPSS, there also exist a number of recently developed functionals that perform comparable to or worse than LSDA, indicating very uneven progress along the axis of improving densities.}

On the other hand, one encouraging note is the fact that the mBEEF mGGA is in fact the \textit{best} semi-local functional tested (with SCAN not being very far behind, despite not being fitted to bonded systems), so there are grounds for optimism that future hybrid mGGAs can also do better for predicting ${\mu}$ than the best present hybrids.
The ability of some mGGAs to eliminate spurious partial charges during dissociation of FH and FCl is another hopeful feature.
Thus future DFA development, particularly involving mGGAs and hybrid mGGAs, could likely benefit from fitting to, or testing on $\vec{\mu}$ in order to seek further improvements, and avoid major failures (e.g. the case of  MN12-L\cite{mn12l} for NaLi). 
We hope that our dataset of computed benchmark dipole moments will assist in development and/or testing of future functionals that improve energetics and also predict highly accurate densities (by contrast, direct comparison against experimental ${\mu}$ values folds in errors due to non-electronic effects such as vibrational averaging). This would take us closer to the exact functional, and in the process assist scientists who are investigating systems where energetics alone are not the sole property of interest. It may likewise be valuable to develop reliable benchmark databases of computed responses of the electron density to perturbations, such as polarizabilities. {Previous studies have assessed the performance of DFAs in reproducing experimental polarizabilities\cite{thakkar2015well,hickey2014benchmarking}, but the presence of nuclear quantum effects necessitates the development of a database of benchmark equilibrium polarizabilities, which so far does not appear to exist}.

\section*{Supporting Information}
Geometries, benchmark dipole moments, table of errors and basis set convergence. 
\section*{Acknowledgement}
D.H. would like to thank Dr Susi Lehtola for helpful discussions. D.H. was funded by a Berkeley Fellowship. This research was supported by the Director, Office of Science, Office of Basic Energy Sciences, of the U.S. Department of Energy under Contract No. DE-AC02-05CH11231.
\bibliography{references}
\end{document}